\documentclass{article}
\usepackage{arxiv}
\usepackage{comment}
\usepackage[utf8]{inputenc} % allow utf-8 input
\usepackage[T1]{fontenc}    % use 8-bit T1 fonts
\usepackage{hyperref}       % hyperlinks
\usepackage{url}            % simple URL typesetting
\usepackage{booktabs}       % professional-quality tables
\usepackage{amsfonts}       % blackboard math symbols
\usepackage{nicefrac}       % compact symbols for 1/2, etc.
\usepackage{microtype}      % microtypography
\usepackage{lipsum}
\usepackage{graphicx}
\usepackage{amsmath}
\usepackage{amssymb}
\usepackage{xcolor}
\usepackage{graphicx}    % for \includegraphics
\usepackage{subcaption}  % for subfigures
\usepackage{longtable}
\usepackage{array}
\usepackage{booktabs}
\usepackage{rotating}
\usepackage{multirow}
\usepackage{enumitem}
\usepackage{tikz}
\usetikzlibrary{positioning}
\usetikzlibrary{arrows.meta}
\usetikzlibrary{shapes.geometric}
\usetikzlibrary{calc}
\graphicspath{ {./images/} }

\title{ARIA: Adaptive Retrieval Intelligence Assistant - A Multimodal RAG Framework for Domain-Specific Engineering Education}

\author{
 Yue  Luo \thanks{Authors contributed equally to this work.}\\
  Engineering Mechanics\\
  Dalian University of Technology\\
  Dalian, Liao 116024 \\
  \texttt{yueluo@mail.dlut.edu.cn} \\
   \And
 Dibakar Roy Sarkar \footnotemark[1]\\
  Civil and Systems Engineering\\
  Johns Hopkins University\\
  Baltimore, MD 21218 \\
  \texttt{droysar1@jh.edu} \\
  \And
 Rachel Herring Sangree\\
  Civil and Systems Engineering\\
  Johns Hopkins University\\
  Baltimore, MD 21218 \\
  \texttt{sangree@jhu.edu} \\
  \And
 Somdatta Goswami\thanks{Corresponding author: \texttt{somdatta@jhu.edu}} \\
  Civil and Systems Engineering\\
  Johns Hopkins University\\
  Baltimore, MD 21218 \\
  \texttt{somdatta@jhu.edu} \\
}

\begin{document}
\maketitle
\begin{abstract}
Developing effective, domain-specific educational support systems is central to advancing AI in education. Although large language models (LLMs) demonstrate remarkable capabilities across diverse domains, they face significant limitations in specialized educational applications, including hallucinations, a limited ability to update or expand their knowledge base, and a lack of domain-specific expertise. Current approaches like fine-tuning require complete model retraining for updates, creating substantial computational overhead, while general-purpose LLMs often provide inaccurate responses in specialized academic contexts. We argue that these limitations arise from their reliance on generalized training data, which fails to exploit the structured, course-specific knowledge essential for effective educational guidance. To address this, we propose ARIA (Adaptive Retrieval Intelligence Assistant) as a Retrieval-Augmented Generation (RAG) framework for creating intelligent teaching assistants across diverse university-level courses. ARIA leverages a multimodal content extraction pipeline combining Docling for structured document analysis, Nougat for mathematical formula recognition, and the GPT-4 Vision API for engineering diagram interpretation. It further incorporates the open-source \texttt{e5-large-v2} embedding model, selected for its high semantic performance and low query latency. This design enables accurate processing of complex educational materials while maintaining pedagogical consistency through carefully engineered prompts and response control mechanisms. We evaluate ARIA using lecture material from \textit{Statics and Mechanics of Materials}, a sophomore-level civil engineering course at Johns Hopkins University, benchmarking performance against ChatGPT-5 using multiple educational-effectiveness criteria. Results demonstrate that ARIA achieves 97.5\% accuracy in domain-specific question filtering and superior pedagogical performance in conceptual explanation quality and structured problem-solving approaches. ARIA correctly answered all 20 relevant course questions while appropriately rejecting 58 of 60 non-relevant queries, achieving 90.9\% precision and 100\% recall. Response quality averaged 4.89/5.0 across all categories. These findings provide evidence that ARIA's course-agnostic architecture and RAG-enhanced knowledge base represent a suitable framework for scalable, domain-specific educational AI deployment. The ARIA web application is available at \url{https://aria-static-mechanics-ta.streamlit.app/}.
\end{abstract}

% keywords can be removed
\keywords{Retrieval-Augmented Generation \and Educational AI \and Intelligent Teaching Assistant \and Multimodal Content Extraction \and Engineering Education \and Large Language Models \and Domain-Specific AI \and STEM Learning Support}

\section{Introduction}

Large Language Models (LLMs) have demonstrated remarkable capabilities across diverse domains, answering complex questions and providing solutions that suggest sophisticated reasoning abilities \cite{brown2020language, achiam2023gpt, touvron2023llama}. Despite their impressive performance, LLMs face significant limitations in specialized applications, particularly the phenomenon of hallucination, where models generate factually incorrect responses with apparent confidence \cite{ji2023survey, huang2025survey}. This limitation stems primarily from their training on vast, generalized datasets that provide broad knowledge but lack domain-specific expertise \cite{gao2023retrieval, goldberg2016primer}. Consequently, while LLMs excel at general tasks, their accuracy diminishes when applied to specialized or knowledge-intensive domains that require precise, current information \cite{petroni2019language,roberts2020much}. These limitations make current LLMs unreliable as course-specific teaching assistants, particularly in engineering domains where interpretation of formulas, diagrams, and step-by-step methods is essential.

Two primary methodologies have emerged to address these limitations and enable domain-specific adaptation of LLMs: fine-tuning and Retrieval-Augmented Generation (RAG) \cite{lewis2020retrieval, gao2023retrieval, qin2024tool}. Fine-tuning allows organizations to adapt existing LLMs to their specific information needs by training models on proprietary or domain-specific datasets, producing commendable performance in specialized contexts \cite{devlin2019bert, kenton2019bert, shojaei2025ai}. However, this approach requires complete model retraining whenever new information becomes available or existing documents undergo modifications, creating significant computational overhead and maintenance challenges \cite{goldberg2016primer, ovadia2023fine}. RAG offers an alternative solution by dynamically retrieving relevant information from external knowledge bases and incorporating this context into the generation process \cite{lewis2020retrieval, shuster2021retrieval}. While RAG systems may not achieve the same level of domain-specific performance as fine-tuned models, they provide considerable advantages in maintaining current information, offering interpretability through traceable source references, and enabling continuous knowledge updates without retraining \cite{guu2020retrieval, borgeaud2022improving, chen2023walking}.

% paragraph to flow to why ARIA?
These developments have particular relevance in educational settings, where there is a growing demand for scalable, accurate, and interactive teaching aids that support educators while ensuring student privacy and equitable access \cite{swacha2025retrieval, wang2024survey}. We introduce ARIA (Adaptive Retrieval Intelligence Assistant), a general RAG-based framework designed to create intelligent teaching assistants for university-level courses across diverse academic domains. The framework's modular architecture enables rapid specialization to any course by incorporating specific course materials as the knowledge base, transforming the system into a domain-expert virtual teaching assistant. Unlike generic educational chatbots, ARIA provides course-specific guidance tailored to the instructor's teaching style and course content, helping students navigate complex concepts when human teaching assistants are unavailable. The framework incorporates configurable restrictions to ensure responses remain relevant to the designated subject matter, preventing off-topic interactions while maintaining academic focus. This approach addresses the substantial workload challenges faced by teaching assistants in large courses while providing students with consistent, high-quality academic support available around the clock.

To demonstrate the framework's effectiveness, we implemented ARIA using course materials from Statics and Mechanics of Materials, though the methodology applies universally to any academic discipline. To guide this investigation, we focus on the following research questions:\vspace{-2pt}
\begin{enumerate}[nosep]
    \item [\textbf{RQ1:}] Can a multimodal, course-specific RAG framework reduce hallucinations and improve domain specificity and response reliability in STEM educational settings?
    \item [\textbf{RQ2:}] Does incorporating structured multimodal extraction (text, formulas, diagrams) improve retrieval quality and downstream pedagogical accuracy compared to general-purpose LLMs?
    \item [\textbf{RQ3:}] Can a course-agnostic architecture be rapidly adapted to new subjects while preserving reliable subject-specific response filtering?
\end{enumerate}\vspace{-4pt}
These questions provide the foundation for ARIA’s design and evaluation. The system employs a sophisticated multimodal content extraction pipeline that integrates three complementary technologies: Docling for structured document analysis and table extraction, Nougat (facebook/nougat-base) for precise mathematical formula and equation recognition \cite{blecher2023nougat}, and GPT-4o Vision API for comprehensive analysis of engineering diagrams, free body diagrams, structural schematics, and visual content. 

This three-pronged approach processes PDF or docx course materials by converting each page to high-resolution images, systematically extracting text structures and mathematical expressions, and providing detailed descriptions of visual elements including engineering concepts and problem-solving methodologies. The multimodal extractor generates comprehensive knowledge representations that capture both symbolic mathematical content and contextual engineering information, which are subsequently converted to vector embeddings using Sentence Transformers and stored in pickle for efficient retrieval during query processing \cite{reimers2019sentence}. The framework's course-agnostic design ensures that specialization for new domains requires only substituting the source materials and minimal modifications to the extraction prompts, making it readily adaptable to courses ranging from engineering and sciences to humanities and social sciences.

ARIA represents a significant advancement in AI-assisted education, offering a scalable solution that enhances learning efficiency and student engagement across diverse academic disciplines through intelligent, context-aware interactions. Unlike prior educational LLM frameworks, ARIA introduces the first multimodal extraction pipeline combining Docling, Nougat, and GPT-4 Vision to create a course-specific RAG system capable of processing diagrams, formulas, and structured content at scale. The framework's contributions to educational technology can be summarized as follows:
\begin{enumerate}[leftmargin=*, nosep]
    \item A comprehensive, course-agnostic RAG-based workflow that can be rapidly specialized for any university course by incorporating domain-specific materials while maintaining strict academic focus.
    \item A robust multimodal embedding system capable of accurately extracting and processing mathematical equations, scientific diagrams, and structured data across disciplines.
    \item Intelligent response control mechanisms that ensure subject-specific interactions while providing educational guidance rather than direct problem solutions, adaptable to any course's pedagogical requirements.
    \item A prototype web application that seamlessly integrates AI-generated responses with traceable references to original course materials, demonstrable across different academic domains.
\end{enumerate}

The remainder of the paper is organized as follows. Section 2 reviews related work, including large language models, their applications in education, and their current limitations. Section 3 details the methodology underlying ARIA’s workflow, the architecture of the RAG-based knowledge repository, and the implementation of the web interface. Section 4 presents experimental results examining data extraction performance, embedding model efficiency, and a comparative evaluation of ARIA against general-purpose LLMs in terms of educational effectiveness. Section 5 provides concluding remarks and outlines directions for future research.

\section{Related Work}

\subsection{Large Language Models and Domain Adaptation}

The advent of transformer architectures and attention mechanisms revolutionized natural language processing, establishing the foundation for modern LLMs \cite{vaswani2017attention}. This breakthrough enabled sophisticated models including BERT \cite{devlin2019bert}, ChatGPT \cite{schulman2022chatgpt}, LLaMA \cite{touvron2023llama}, and DeepSeek \cite{bi2024deepseek}. However, domain-specific adaptation remains computationally expensive and resource-intensive. Parameter-efficient fine-tuning techniques have emerged to address these limitations. Low-Rank Adaptation (LoRA) exemplifies this approach by introducing trainable low-rank matrices into frozen transformer layers, enabling effective specialization with minimal parameter updates \cite{hu2022lora}. This methodology leverages insights about neural networks' intrinsic low-dimensional structure \cite{li2018measuring,aghajanyan2020intrinsic}.

Alternatively, Retrieval-Augmented Generation (RAG) provides external knowledge integration without parameter modifications \cite{lewis2020retrieval}. RAG systems encode user queries through embedding models, retrieve semantically similar documents from pre-indexed knowledge bases, and augment the original query with retrieved context for enhanced generation. This approach offers distinct advantages for maintaining current information and ensuring response traceability to source materials.

\subsection{Educational Applications of Large Language Models}

The integration of generative artificial intelligence in academic environments has witnessed substantial growth, manifesting across diverse educational contexts \cite{xu2024leveraging, wang2024large}. Notable implementations include code debugging assistance for programming students \cite{yang2024debugging}, virtual classroom simulation platforms \cite{zhang2024simulating}, and intelligent tutoring systems \cite{hicke2023ai, mehta2023can}.A particularly relevant contribution comes from Faghih Shojaei et al. (2025) \cite{shojaei2025ai}, who developed AI-University (AI-U), an educational framework combining LoRA-based LLaMA-3.2 fine-tuning with RAG synthesis for graduate finite element method instruction. Their approach utilized Mathpix for LaTeX conversion of 4,648 systematically generated question-answer pairs derived from instructional materials.

Our methodology diverges significantly from this foundation. Rather than relying on external conversion tools, we implement a comprehensive multimodal extraction pipeline capable of directly processing diverse content types including text, code, tables, images, mathematical formulas, and handwritten annotations from authentic course materials. Additionally, our framework incorporates instructor-provided question-answer pairs and implements specialized tools that operate independently of LLM generation, enabling dynamic answer validation and standardized response control mechanisms that enhance educational reliability beyond purely generative approaches.

% \paragraph 3 shortcoming and concerns:\\

\subsection{Limitations and Implementation Challenges}

Despite significant potential, several barriers impede widespread LLM adoption in educational settings. Privacy and security concerns emerge when utilizing commercial platforms, particularly regarding sensitive student information and proprietary academic materials  \cite{chan2023comprehensive}. Educational equity represents another critical challenge, as students from privileged backgrounds demonstrate greater benefits from AI literacy and access \cite{yu2024whose}.

Current RAG implementations face substantial technical constraints. While context window limitations have improved with recent LLMs, they remain inadequate for processing extensive reference corpora effectively. Knowledge graph integration shows promise for enhancing RAG scalability and performance \cite{xing2024retrieval}. Additionally, RAG systems encounter fundamental retrieval and generation failures that directly impact response quality, particularly when relevant documents are not retrieved or when generative modules fail to synthesize retrieved information effectively.

STEM education presents unique multimodal challenges requiring sophisticated embedding models capable of processing mathematical formulas, diagrams, and handwritten annotations. The embedding process itself offers potential for developing more cost-effective, domain-specific models trained on physics content, potentially reducing dependency on expensive commercial solutions. However, standardized evaluation frameworks for assessing physics-specific LLM capabilities remain absent. Ethical considerations necessitate rigorous accuracy verification, as incorrect responses could fundamentally compromise student learning outcomes rather than enhance educational experiences.

\section{Methodology}

\subsection{System Overview}

\begin{figure}[!tbh]
    \centering
    \includegraphics[width=0.95\textwidth]{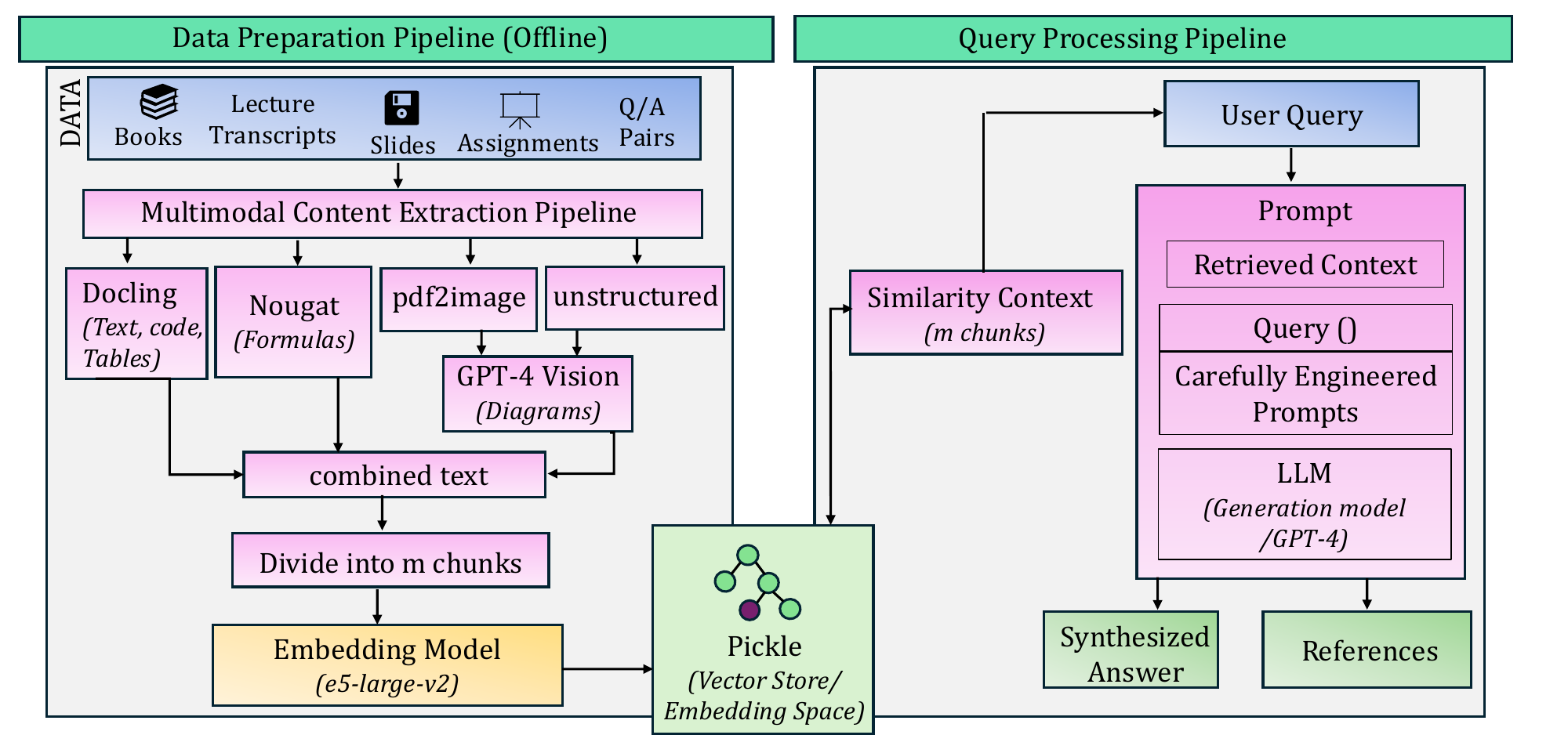} 
    \caption{Architecture of the ARIA web application demonstrating the offline data preparation pipeline (left) and real-time query processing pipeline (right). The system processes multimodal educational content (books, lecture transcripts, slides, assignments, Q\&A pairs) through specialized extraction modules including Docling for text and tables, Nougat for mathematical formulas, and GPT-4 Vision for diagrams. Content is segmented into chunks and embedded using text-embedding3-large, with vectors stored in a Pickle-based database. During query processing, the system performs similarity-based retrieval of relevant chunks, combines them with user queries in carefully engineered prompts, and generates responses using GPT-4 as the synthesis model, providing both answers and numbered source references for traceability.}
    \label{fig:schematic_ARIA}
\end{figure}

The ARIA (Adaptive Retrieval Intelligence Assistant) framework is a Retrieval Augmented Generation-based learning assistant designed to deliver context-grounded and traceable educational support. The system architecture employs a multi stage pipeline that combines offline preparation of course materials with real time semantic retrieval and synthesis. As illustrated in Figure~\ref{fig:schematic_ARIA}, student queries are processed through similarity based search, contextual retrieval, and large language model based response generation with citation linkage to specific instructional materials \cite{lewis2020retrieval}. These responses are further shaped by carefully designed pedagogical guardrails that regulate domain fidelity, instructional coherence, and the overall quality of generated explanations.

The interaction begins when a student submits a query through the web interface. The system identifies relevant excerpts of course content from the embedded knowledge base, integrates retrieved context with the question through prompt conditioning, and generates explanations that include numbered references for verification and deeper learning. The system maintains conversational continuity to support multi turn exchanges and adapts explanation depth according to the student’s progression.

\subsection{Multimodal Extraction Pipeline}

The training dataset comprises course materials from the Statics and Mechanics of Materials course (EN.560.201) offered in the Department of Civil and Systems Engineering Department at Johns Hopkins University. The material spans core topics of forces, moments, stress strain relationships, deformation analysis, and equilibrium conditions, delivered through 33 slide-based lectures with voice-over instruction and 58 curated question answer pairs derived from 22 assignments \cite{beer2006mechanics}.

\begin{figure}[!tbh]
    \centering
    \includegraphics[width=0.95\textwidth]{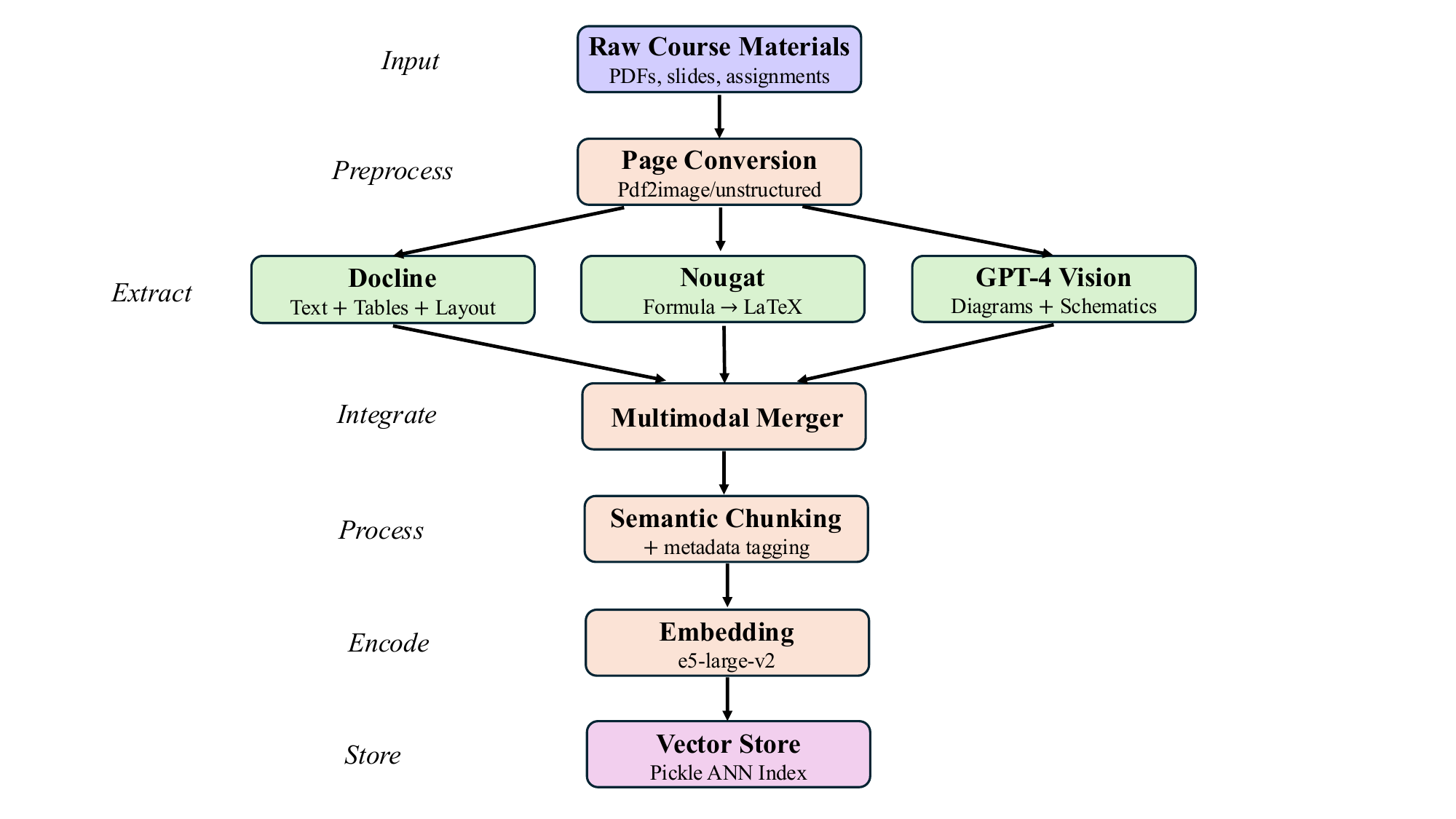} 
    \caption{Multimodal content extraction pipeline for ARIA. Course materials are converted to pdf or images and processed through three parallel extractors (Docling, Nougat, GPT-4 Vision), merged into unified representations, semantically chunked with metadata, embedded, and indexed for retrieval. }
    \label{fig:multimodal_pipeline}
\end{figure}

The content extraction pipeline follows a structured multimodal workflow that corresponds to the stages shown in Figure \ref{fig:multimodal_pipeline}. The process begins with raw instructional materials such as PDFs, slides, and assignments. These documents are first converted into page level image or pdf representations during the preprocessing stage. The images or pdf then enter three parallel extraction pathways that specialize in different modalities of engineering course content.

Docling functions as the primary text understanding engine. It uses DocLayNet for layout analysis and TableFormer for table structure recognition, which allows precise extraction of textual descriptions, tabular information, and document layout \cite{livathinos2025docling}. In parallel, the Nougat Vision Encoder Decoder Model captures mathematical expressions and converts them into structured LaTeX, ensuring high fidelity preservation of equations across different formatting styles. A third pathway powered by GPT 4 Vision identifies and interprets visual elements that are essential in mechanics education, including engineering diagrams, free body sketches, schematics, and problem solution workflows.

As illustrated in Figure \ref{fig:multimodal_pipeline}, the outputs from these three extractors are then integrated into a unified multimodal representation. This integration step consolidates text, formulas, visual annotations, and structural layout information. The merged content is processed through a semantic chunking stage with metadata tagging that preserves conceptual boundaries and instructional relationships across topics. These carefully prepared chunks are then encoded using the text embedding 3 large model to produce high dimensional vector representations suitable for retrieval.

The resulting embeddings are stored in a pickle based approximate nearest neighbor index, which forms the vector store used throughout the ARIA retrieval augmented generation pipeline. This design ensures that textual, mathematical, and visual components of the course are accurately captured, searchable, and ready for downstream educational use.

\subsection{Embedding and Retrieval Layer}

OpenAI’s text-embedding-3-large model encodes input text into dense numerical vectors and, by default, produces 3072-dimensional embeddings that capture semantic relationships in engineering oriented content. It maps conceptually similar passages to nearby positions in vector space and offers an adjustable dimensions parameter to balance representational capacity with efficiency. ARIA uses this embedding model to index educational material by converting each concept-aligned chunk into a vector that reflects its meaning and mathematical relevance.

ARIA segments course materials using a hierarchical chunking strategy aligned with topic boundaries and cognitive structure, and each chunk includes metadata descriptors such as topic domain, source document reference, difficulty tier, and prerequisite knowledge. These embeddings, together with their metadata, are stored in a vector index and support similarity search at query time. When a user submits a question, ARIA embeds the query and performs approximate nearest neighbor retrieval to surface the most semantically related chunks; simple queries return tightly scoped foundational material, while multi topic questions expand retrieval across several conceptual domains.

A pickle based vector storage and approximate nearest neighbor indexing support low latency retrieval. The retrieval layer dynamically adjusts based on query category and conceptual complexity, enabling efficient and pedagogically aligned performance as the knowledge base scales.

\subsection{Pedagogy Controls and Response Generation}

\begin{figure}[!tbh]
    \centering
    \includegraphics[width=\textwidth]{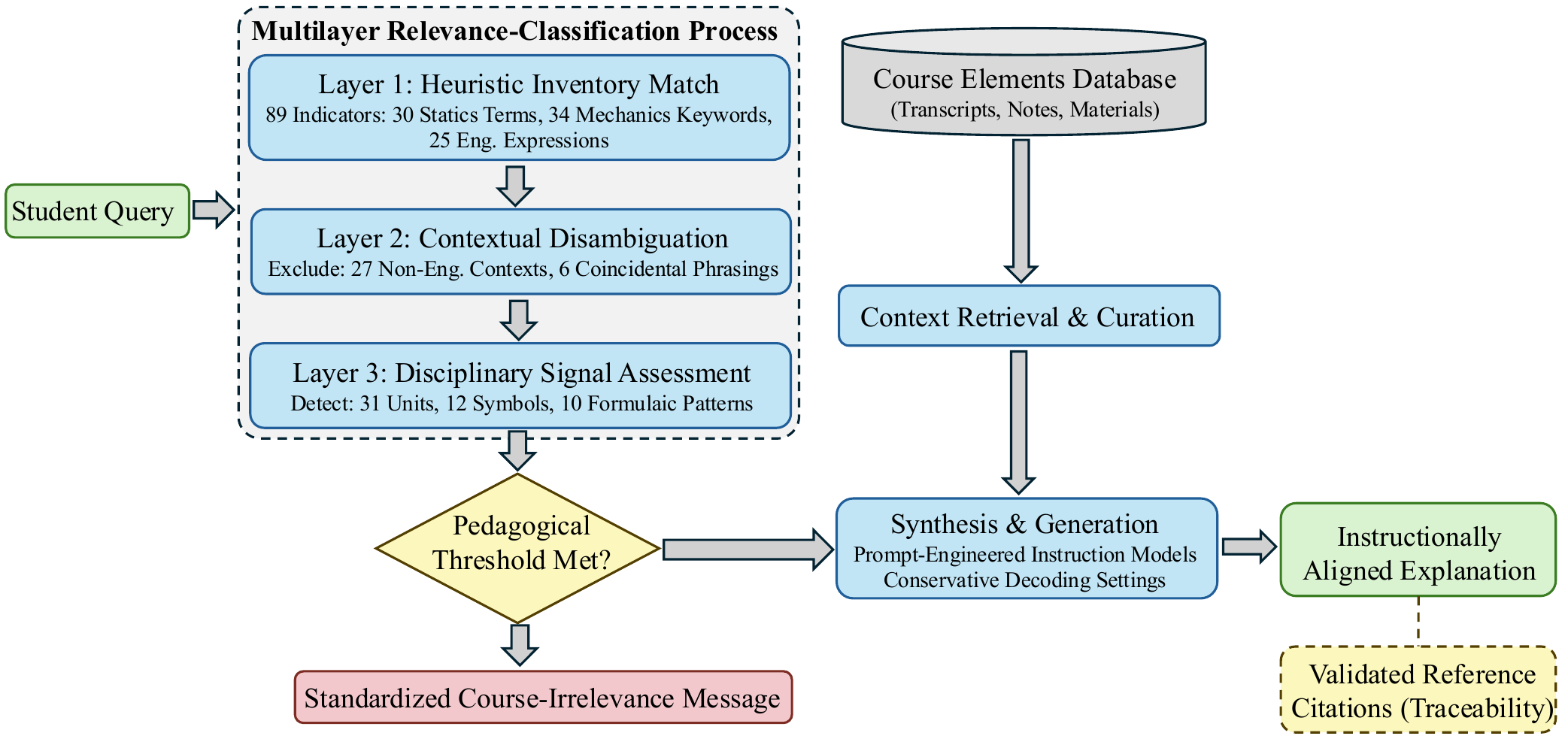} 
    \caption{Schematic representation of the ARIA pedagogical control and response generation framework. The process illustrates the multilayer relevance classification, decision thresholds, and the citation-validated generation pipeline.}
    \label{fig:gaurdrail}
\end{figure}

The ARIA framework employs a structured set of pedagogical control mechanisms to ensure that all generated responses remain strictly confined to the Statics and Mechanics of Materials curriculum as shown in Figure \ref{fig:gaurdrail}. Central to this framework is a multilayer relevance-classification process that evaluates whether a student’s question aligns with course concepts. This process draws on an extensive heuristic inventory composed of 30 Statics-related keyboards, 34 Mechanics-related keywords, and 25 additional engineering-related keywords, forming a combined domain vocabulary of 89 indicators. These lexical cues allow the system to recognize legitimate course-related inquiries while maintaining high precision in topic adherence.

To avoid misclassification, ARIA incorporates a contextual disambiguation layer that excludes 27 common non-engineering contexts and identifies 6 categories of coincidental phrasing that may mimic technical terminology. Additional disciplinary signals are assessed through the detection of 31 engineering measurement units, 12 mathematical or mechanical symbols, and 10 formulaic patterns characteristic of equilibrium analysis and stress-strain relationships. Questions failing to meet these pedagogical thresholds receive a standardized course-irrelevance message, thereby preventing unsupported or tangential discourse.

For queries deemed relevant, the system synthesizes retrieved context and the student query through prompt-engineered instruction models to produce explanations accompanied by reference citations that link directly to course elements including transcripts, prepared notes, or instructional materials. This architecture enables traceability, supports source verification, and reinforces evidence-based learning.

Response generation operates under conservative decoding settings to minimize stylistic drift and maintain terminological precision. Retrieved materials are curated to ensure conceptual accuracy and instructional coherence, and citations are appended only when validated course sources are available. Collectively, these pedagogical controls ensure that ARIA functions as a disciplined, domain-bounded teaching assistant capable of delivering accurate, verifiable, and instructionally aligned explanations while preventing the system from extending beyond the authorized curriculum.

\subsection{Web Interface}

\begin{figure}[h!]
    \centering
    \includegraphics[width=0.95\textwidth]{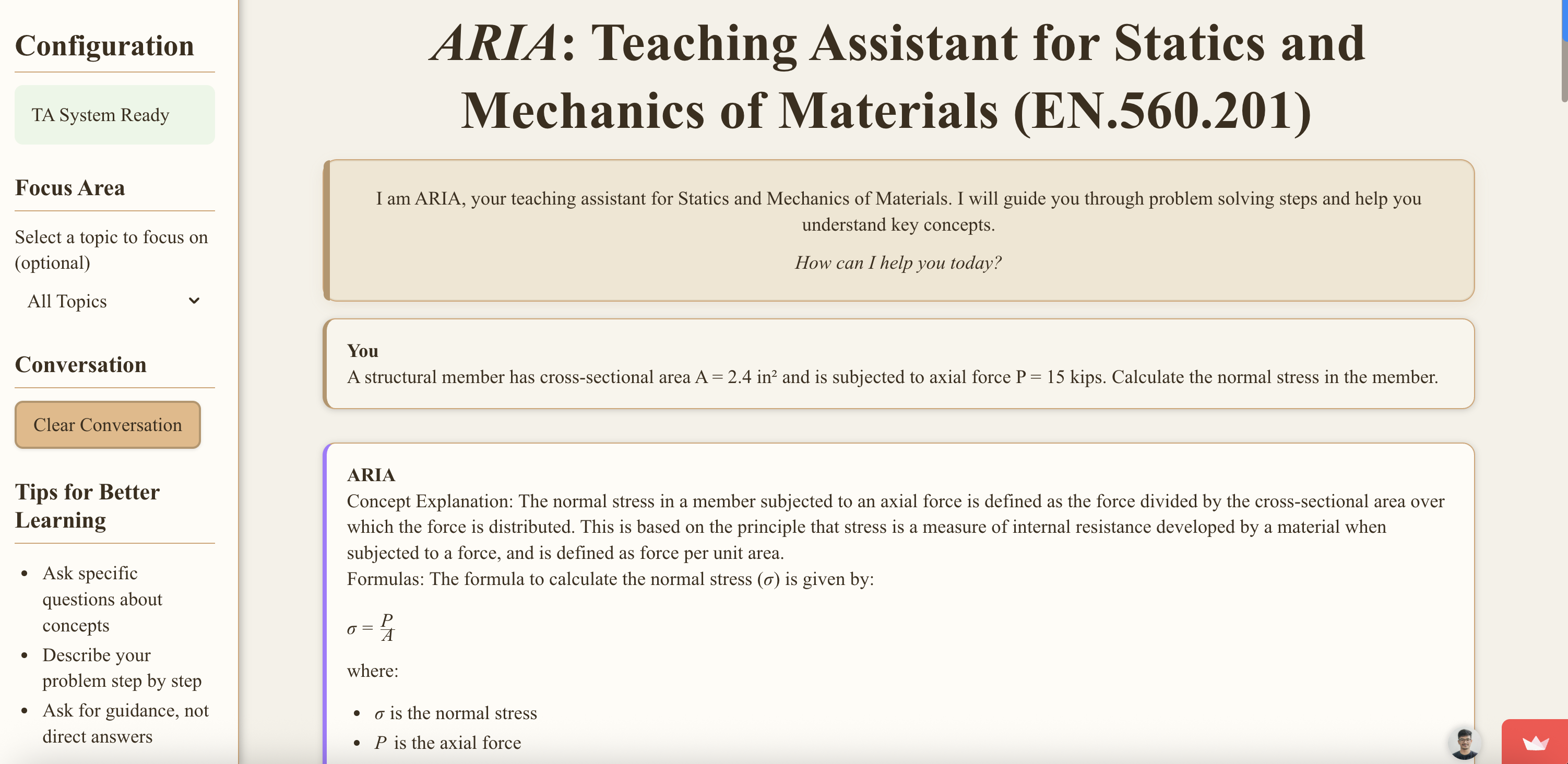} % replace with your file name
    \caption{Demonstration of the web application, available at \url{https://aria-static-mechanics-ta.streamlit.app/}}
    \label{fig:web_app}
\end{figure}

The ARIA web application is implemented using Streamlit and is accessible at \url{https://aria-static-mechanics-ta.streamlit.app/}. Figure \ref{fig:web_app} shows the interface layout. The design emphasizes clarity and accessibility, supported by custom CSS with a muted academic oriented color palette. The interface includes real time feedback components and topic level filtering to support personalized study workflows.

The backend integrates securely with OpenAI language models and includes rate limiting, API key management, and error handling protocols to maintain reliable operation. Execution processes are optimized to support consistent educational usage and query throughput.

\section{Experiment}
            
To establish the necessity and effectiveness of ARIA’s multimodal extraction pipeline, the experimental analysis began with a component-wise ablation study comparing individual extraction tools against the integrated system. A six-page lecture slide deck on torsion analysis in cylindrical shafts from the Statics and Mechanics curriculum was selected as the source document because it contains mathematical formulas, engineering diagrams, and explanatory text, making it an appropriate test of multimodal capability. Each extraction method processed the identical PDF independently. PyPDF2 served as a baseline text extractor, Nougat provided mathematical formula recognition, and GPT-4 Vision interpreted diagrammatic and visual content. The integrated ARIA configuration combined Docling for text and structural parsing, Nougat for symbolic mathematical extraction, and GPT-4 Vision for visual analysis, with all outputs merged into unified multimodal representations. This design enabled direct comparison between isolated tools and the full pipeline to determine the contribution and necessity of each component.

Building on this foundation, the broader experimental evaluation assessed ARIA as a specialized RAG-enhanced educational AI system built upon the GPT-4 architecture. Its configuration included a maximum response length of 400 tokens, a temperature of 0.7, and presence and frequency penalties of 0.1. ARIA was compared against ChatGPT-5 accessed through the standard web interface, which dynamically employs both GPT-5 and GPT-5 Mini models. The evaluation included two complementary analyses: the first examined ARIA’s ability to filter and retain domain-specific questions to maintain conceptual focus within statics and mechanics; the second assessed educational effectiveness to determine whether ARIA’s tailored architecture and curated knowledge base provide learning advantages over general-purpose models.

\subsection{Multimodal Extraction Pipeline Validation}

We measured extraction performance by total character count across the six lecture slides on torsion analysis in cylindrical shafts. This document represents typical engineering course materials containing mathematical formulas, engineering diagrams, and explanatory text. Table~\ref{tab:extraction_comparison} presents comparative results for each extraction method.

\begin{table}[h!]
\centering
\caption{\textbf{Extraction Performance Comparison Across Methods.} Comparative assessment of five extraction methods applied to six engineering lecture slides on torsion analysis. Total Characters quantifies extraction volume. Text Content categorizes textual extraction quality: None (no text), Partial (incomplete), OCR-based (image-to-text conversion), Structured (hierarchical formatting preserved), or Complete (full content with formatting). Formula Content assesses mathematical formula handling: None (not extracted), Placeholders (position marked, content unprocessed), or LaTeX format (equations converted to structured notation). Diagram Content evaluates visual element preservation: None (diagrams missing) or Detailed (comprehensive diagram information captured).}
\label{tab:extraction_comparison}
\begin{tabular}{lcccc}
\toprule
\textbf{Extraction Method} 
& \textbf{Total Characters} 
& \textbf{Text Content} 
& \textbf{Formula Content} 
& \textbf{Diagram Content} \\
\midrule

PyPDF2 
& 0 
& None 
& None 
& None \\

Docling 
& 3{,}924 
& Structured 
& Placeholders 
& None \\

Nougat 
& 4{,}268 
& Partial 
& LaTeX format 
& None \\

GPT-4 Vision 
& 12{,}568 
& OCR-based 
& Natural language 
& Detailed \\

\textbf{ARIA-Full} 
& \textbf{20{,}760} 
& \textbf{Complete} 
& \textbf{LaTeX format} 
& \textbf{Detailed} \\

\bottomrule
\end{tabular}
\end{table}

PyPDF2 achieved zero extraction on image-based PDFs, confirming the incompatibility of standard libraries with modern educational materials that employ image rendering for visual quality. Docling extracted 3{,}924 characters (18.9\% of total), successfully providing structured text with preserved document hierarchy but marking formulas as unprocessable placeholders (\texttt{<!-- formula-not-decoded -->}). Nougat extracted 4{,}268 characters (20.6\%), successfully converting all eight mathematical formulas into proper LaTeX format while providing only fragmented contextual text surrounding equations.

GPT-4~Vision extracted 12{,}568 characters (60.5\%), offering comprehensive diagram descriptions including dimensional specifications and geometric relationships, along with complete text through optical character recognition. However, mathematical expressions were rendered in natural language descriptions rather than structured LaTeX format, preventing computational manipulation.

ARIA-Full which is combination of all of them extracted 20{,}760 characters, representing complete information capture and a 65.2\% increase over the best individual method. This integration captures distinct, non-redundant information types: structured text with preserved document hierarchy (Docling), computationally usable LaTeX formulas enabling symbolic manipulation (Nougat), and detailed visual semantics including engineering diagram analysis (GPT-4~Vision). Each component addresses a unique content modality essential for engineering education. The results demonstrate that comprehensive educational material processing requires orchestrated multimodal extraction, as no single tool provides sufficient coverage across text, mathematical, and visual content dimensions.

\subsection{Embedding Model Performance Evaluation}

To assess whether embedding model selection affects retrieval quality for engineering education content, we conducted a systematic comparison of five embedding models spanning different dimensionalities and architectural approaches. The evaluated models included \texttt{text-embedding-3-large} (3072 dimensions), \texttt{text-embedding-3-small} (1536 dimensions), \texttt{all-MiniLM-L6-v2} (384 dimensions), \texttt{instructor-xl} (768 dimensions), and \texttt{e5-large-v2} (1024 dimensions). All models were evaluated using identical ARIA-Full extracted content. Performance was assessed using five representative queries spanning formula identification, conceptual understanding, and problem-solving scenarios.

\begin{table}[h!]
\centering
\small
\setlength{\tabcolsep}{4pt}  
\caption{Embedding Model Performance Comparison}
\label{tab:embed_performance}
\begin{tabular}{lccccccc}
\toprule
\textbf{Model} 
& \textbf{Dimensions} 
& \textbf{Accuracy@5} 
& \textbf{MRR} 
& \textbf{NDCG@5} 
& \textbf{Avg Similarity} 
& \textbf{Query Latency (s)} 
& \textbf{Storage (KB)} \\
\midrule

\texttt{text-emb-3-large} 
& 3072 & 1.00 & 1.00 & 1.00 & 0.588 & 1.177 & 346 \\

\texttt{text-emb-3-small} 
& 1536 & 1.00 & 1.00 & 1.00 & 0.604 & 0.769 & 190 \\

\texttt{all-MiniLM-L6-v2} 
& 384 & 1.00 & 1.00 & 1.00 & 0.613 & 0.065 & 73 \\

\texttt{instructor-xl} 
& 768 & 1.00 & 1.00 & 1.00 & 0.798 & 0.849 & 112 \\

\textbf{\texttt{e5-large-v2}}
& \textbf{1024} 
& \textbf{1.00} 
& \textbf{1.00} 
& \textbf{1.00} 
& \textbf{0.851} 
& \textbf{0.178} 
& \textbf{138} \\

\bottomrule
\end{tabular}
\end{table}

We evaluated models using standard information retrieval metrics: Accuracy@5 measures whether relevant content appears within the top five retrieved results; Mean Reciprocal Rank (MRR) quantifies the rank position of the first correct retrieval (ranging from 0 to 1, with 1 indicating a rank-1 match); and Normalized Discounted Cumulative Gain (NDCG@5) measures ranking quality by emphasizing higher-ranked items. Average similarity denotes the mean cosine similarity of the top three retrieved chunks, providing an estimate of retrieval confidence.

All models achieved perfect scores (1.00) for Accuracy@5, MRR, and NDCG@5, successfully identifying relevant content within the top five retrieved items for all queries. This uniform performance suggests that for focused technical domains with tightly scoped, well-curated knowledge bases, embedding dimensionality has minimal impact on retrieval correctness. The concentrated thematic structure of engineering materials produces natural semantic clustering, enabling high retrieval accuracy across diverse embedding architectures.

Despite equivalent accuracy, substantial differences emerged in similarity scores and computational efficiency. The open-source models \texttt{e5-large-v2} and \texttt{instructor-xl} produced significantly higher average similarity scores (0.851 and 0.798, respectively) than the OpenAI embeddings (0.588--0.604), indicating stronger semantic alignment with technical engineering content despite lower dimensionality. Higher similarity reflects stronger retrieval confidence, which becomes increasingly important when applying threshold-based filtering or discriminating between closely related concepts.

Query latency varied dramatically across models. \texttt{all-MiniLM-L6-v2} demonstrated the fastest performance, processing queries approximately 18 times faster than \texttt{text-embedding-3-large} (0.065 seconds versus 1.177 seconds). Storage requirements scaled proportionally with embedding dimensionality: \texttt{text-embedding-3-large} required 346 KB, compared to only 73 KB for \texttt{all-MiniLM-L6-v2}, a key factor when indexing larger educational repositories.

These results demonstrate that lightweight open-source embeddings deliver equivalent retrieval accuracy while offering significant advantages in efficiency and semantic coherence. The \texttt{e5-large-v2} model presents the strongest overall balance, achieving the highest similarity score (0.851), moderate query latency (0.178 seconds), and reasonable storage requirements (138 KB). This makes it an optimal choice for deployment in resource-constrained educational environments where both accuracy and operational efficiency are essential.

\subsection{Domain Specificity and Question Filtering}

This experiment evaluated ARIA's ability to accurately distinguish between questions within its designated educational domain and those outside its scope of expertise. The assessment aimed to validate the system's capacity to maintain focused educational support while establishing clear operational boundaries that prevent scope creep.

We developed a comprehensive evaluation framework consisting of 80 strategically curated questions distributed in four distinct categories. Category 1 contained 20 relevant Statics and Mechanics questions derived directly from actual course assignments, covering force analysis, equilibrium calculations, shear force diagrams, bending moments, torsion problems, and material stress analysis. These questions should generate detailed explanations and step-by-step solutions. Category 2 included 20 engineering-adjacent questions from fluid mechanics, thermodynamics, and electrical engineering. Category 3 encompassed 20 academic non-engineering questions spanning calculus, chemistry, and history. Category 4 comprised 20 general personal inquiries about cooking, travel, and lifestyle advice. Categories 2-4 should trigger appropriate rejection responses with standardized messages.

The evaluation methodology employed rigorous performance metrics derived from information retrieval literature \cite{swacha2025retrieval}. Precision measured the ratio of correctly answered Statics and Mechanics questions to all detailed responses (target >90\%). Recall assessed the proportion of relevant Statics and Mechanics questions receiving appropriate responses (target >95\%). The F1 score provided a balanced classification performance measure (target >92\%). In addition, each response received an appropriateness rating on a five-point scale based on domain accuracy, explanation quality, and educational value.

\begin{figure}[h!]
    \centering
    % First subfigure
    \begin{subfigure}[b]{0.45\textwidth}
        \centering
        \includegraphics[width=\textwidth]{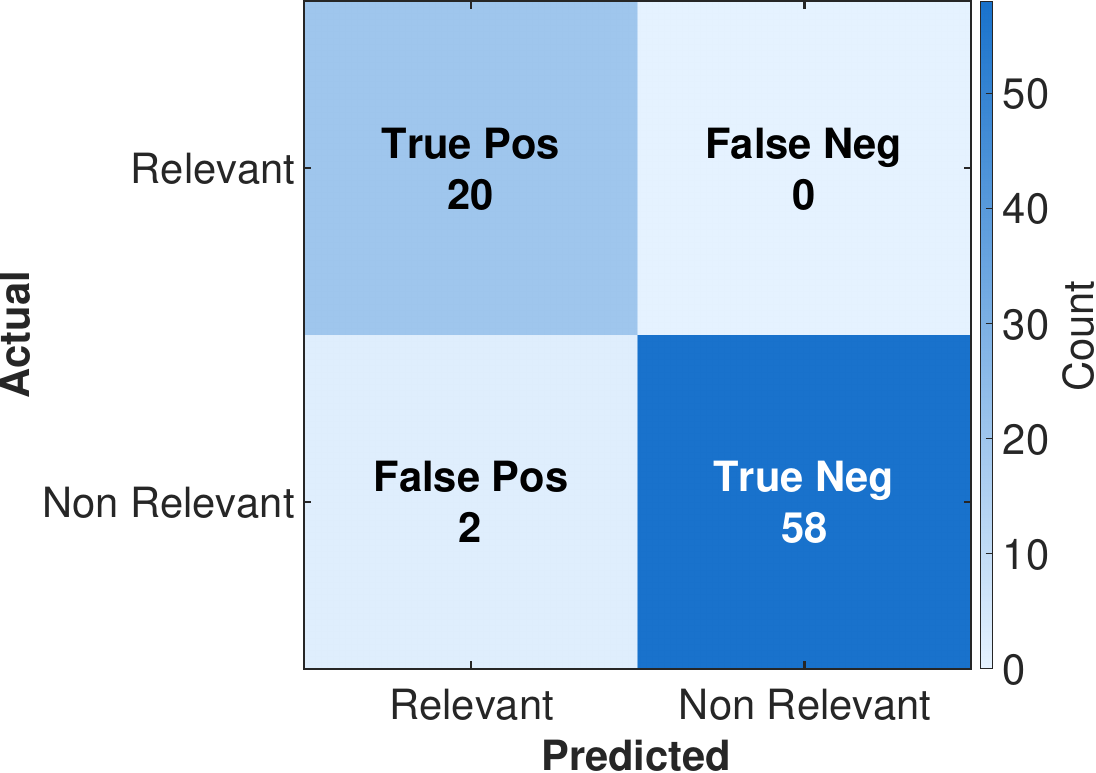}
        \caption{Confusion matrix}
        \label{fig:one}
    \end{subfigure}
    \hfill
    % Second subfigure
    \begin{subfigure}[b]{0.45\textwidth}
        \centering
        \includegraphics[width=\textwidth]{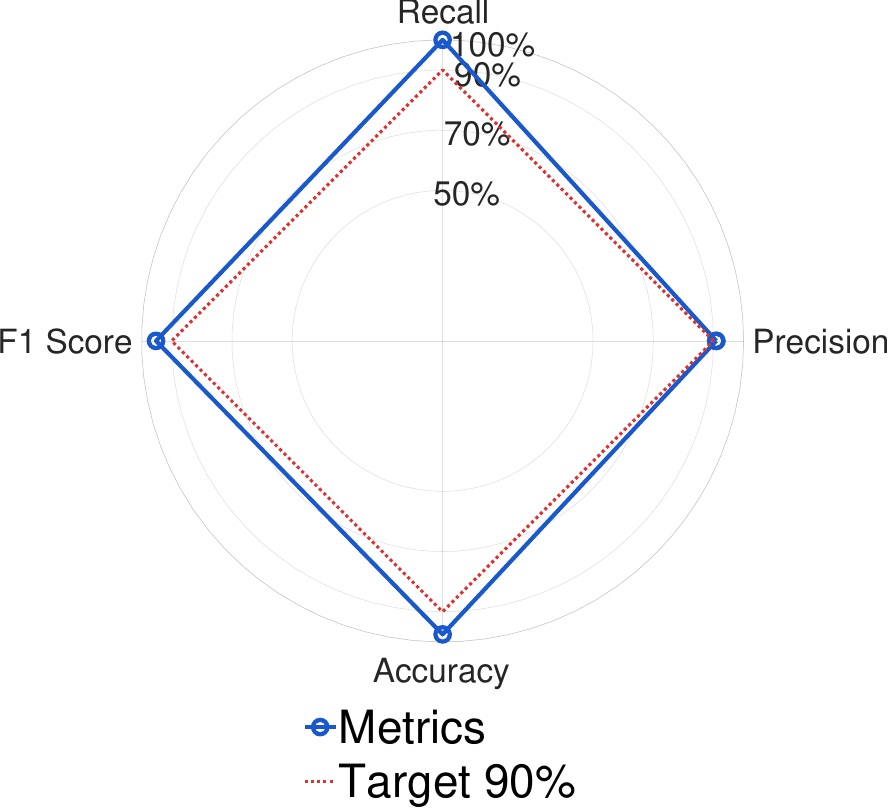}
        \caption{Performance Radar}
        \label{fig:two}
    \end{subfigure}
    
    \caption{Evaluation results of the ARIA system performance on document relevance classification task. The experiment tested the system's ability to distinguish relevant from non-relevant educational documents using a labeled test dataset of 80 samples. Results show (a) confusion matrix with 20 true positives, 0 false negatives, 2 false positives, and 58 true negatives, and (b) performance radar chart displaying key information retrieval metrics (Recall, Precision, F1 Score, and Accuracy) against a 90\% target threshold.}
    \label{fig:exp1_metrics}
\end{figure}

ARIA achieved 90.9\% precision and 100\% recall in domain classification, resulting in an F1-score of 95.2\% and overall accuracy of 97.5\% as showm im Figure \ref{fig:exp1_metrics}. The system correctly answered all 20 relevant statics/mechanics questions while appropriately rejecting 58 of 60 non-relevant queries. Only 2 engineering-adjacent questions were incorrectly answered, likely due to overlapping technical terminology. Response quality averaged 4.89/5.0 across all categories, demonstrating both accurate classification and high pedagogical value.

\subsection{Educational Environment evaluation}

Building upon the validated question set from the domain specificity evaluation, this experiment utilized the similar 20 relevant statics and mechanics questions that successfully demonstrated ARIA's operational competency in the previous assessment. These questions, having been confirmed to fall within ARIA's specialized domain, provided a controlled foundation for comparing educational effectiveness between ARIA's domain-specific architecture and general-purpose AI systems.

The experimental design specifically targeted whether ARIA's RAG-enhanced knowledge base and specialized training translate into superior educational outcomes compared to GPT-5, a state-of-the-art general-purpose language model. By employing the previously validated question set, this approach eliminated potential confounding variables related to domain recognition failures, allowing for a focused comparison of educational quality and pedagogical effectiveness.

Both ARIA and GPT-5 generated responses to each of the 20 questions, which were subsequently evaluated using a streamlined framework of six binary metrics: mathematical correctness, conceptual explanation quality, step-by-step solution clarity, teaching approach versus direct answer provision, question responsiveness, and overall educational value. This binary assessment framework was designed to enable rapid, objective evaluation while minimizing evaluator bias and ensuring consistent scoring criteria across all responses.

\begin{figure}[h!]
    \centering
    % First subfigure
    \begin{subfigure}[b]{0.7\textwidth}
        \centering
        \includegraphics[width=\textwidth]{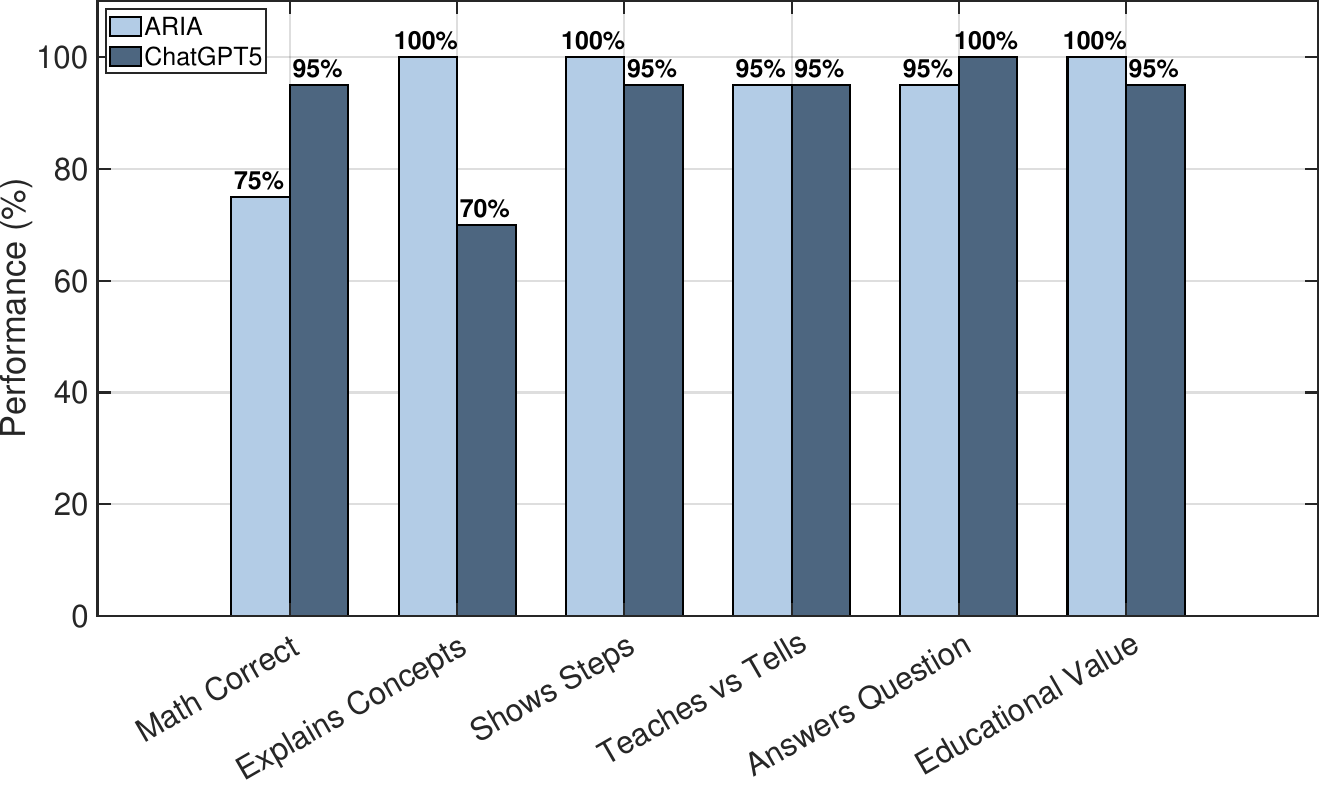}
        \caption{ARIA vs ChatGPT-5: Performance Across Metrics}
        \label{fig2:one}
    \end{subfigure}
    \hfill
    % Second subfigure
    \begin{subfigure}[b]{0.27\textwidth}
        \centering
        \includegraphics[width=\textwidth]{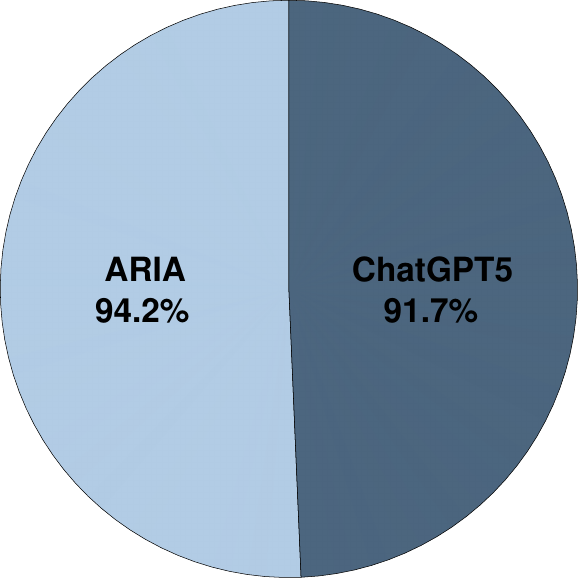}
        \caption{Overall Performance comparison}
        \label{fig2:two}
    \end{subfigure}
    
    \caption{Comparative evaluation of ARIA versus ChatGPT-5 performance on educational quality metrics for Statics \& Mechanics instruction. The experiment assessed both systems across six pedagogical criteria using expert human evaluation, showing (a) performance comparison across specific metrics including mathematical correctness, conceptual explanation ability, step-by-step problem solving, teaching approach quality, question answering accuracy, and overall educational value, and (b) aggregated overall performance scores demonstrating ARIA's 94.2\% versus ChatGPT-5's 91.7\% effectiveness in educational content delivery and student learning support.}
    \label{fig:exp2_metrics}
\end{figure}

As illustrated in Figure \ref{fig:exp2_metrics}, the performance comparison revealed complementary strengths between the two systems. ARIA consistently excelled in pedagogical dimensions, providing structured explanations that emphasized conceptual understanding and systematic problem-solving approaches over direct answer provision. Conversely, GPT-5 demonstrated superior mathematical accuracy and computational precision while maintaining comparable performance across other educational metrics. These findings suggest that ARIA's domain-specific architecture and RAG-enhanced knowledge base effectively optimize for educational quality and conceptual clarity, though mathematical computation represents an area where general-purpose models currently maintain an advantage.

\section{Conclusion}

This paper presents ARIA (Adaptive Retrieval Intelligence Assistant), a comprehensive RAG-based framework designed to address fundamental limitations of general-purpose LLMs in specialized educational contexts. By integrating multimodal content extraction, efficient embedding strategies, and pedagogical control mechanisms, ARIA demonstrates how targeted AI systems can provide superior domain-specific educational support while maintaining strict academic focus and ensuring response traceability.

The framework's primary contribution lies in its validated multimodal content extraction pipeline, which integrates Docling for structured document analysis, Nougat for mathematical formula recognition, and GPT-4 Vision for engineering diagram interpretation. The extraction ablation study demonstrated that comprehensive educational material processing requires orchestrated multimodal extraction rather than reliance on any single tool. While PyPDF2 achieved zero extraction on image-based educational materials, the complete ARIA pipeline captured 20,760 characters, representing a 65.2\% increase over the best individual extractor. This substantial difference reflects distinct, non-redundant information types essential for engineering education support.

The embedding model comparison revealed that lightweight open-source models provide equivalent retrieval accuracy while offering superior efficiency for focused technical domains. All five tested models achieved perfect retrieval metrics, confirming that domain-specific knowledge bases reduce sensitivity to embedding dimensionality. However, e5-large-v2 demonstrated optimal performance balance with highest semantic similarity, moderate latency, and reasonable storage requirements, making it preferable for resource-constrained educational deployments.

ARIA's pedagogical effectiveness was validated through comprehensive evaluation demonstrating 97.5\% accuracy in domain-specific question filtering and superior educational quality compared to ChatGPT-5. The system excelled particularly in pedagogical dimensions, including conceptual explanation quality, structured problem-solving approaches, and teaching-oriented guidance rather than direct answer provision.

Future enhancements include integrating specialized mathematical reasoning modules to address current computational limitations, optimizing document processing architectures to balance accuracy and computational efficiency, as sequential document-by-document analysis yields superior content extraction accuracy compared to batch processing but introduces prohibitive latency, developing scalability solutions for institutional deployment through federated learning approaches, establishing standardized evaluation frameworks assessing learning outcomes and retention, implementing adaptive personalization algorithms based on individual student progress, expanding beyond STEM fields to humanities and social sciences, and incorporating multilingual support for global educational deployment while maintaining data privacy and cross-cultural adaptability.

\section{Acknowledgments}
 The authors would like to acknowledge computing support provided by the Advanced Research Computing at Hopkins (ARCH) core facility at Johns Hopkins University and the Rockfish cluster. ARCH core facility (\url{rockfish.jhu.edu}) is supported by the National Science Foundation (NSF) grant number OAC1920103. The research efforts of DRS and SG are supported by National Science Foundation (NSF) under Grant No. 2436738. Any opinions, findings, conclusions, or recommendations expressed in this material are those of the author(s) and do not necessarily reflect the views of the funding organizations.

\section*{Author contributions}
\noindent Conceptualization: YL, DRS, SG  \\
Investigation: YL, DRS, SG \\
Visualization: YL, DRS, SG  \\
Supervision: SG \\
Writing - original draft: YL, DRS \\
Writing - review \& editing: YL, DRS, SG 

\section*{Data and code availability}
\noindent The codes that support the findings of this study are available on \url{https://github.com/RoyDibs/ARIA_static_mechanics_app}.

\section*{Competing interests}
\noindent The authors declare no competing interest

\section*{Ethics Statement}

\textbf{AI-Generated Responses:} ARIA should be used responsibly. Answers are generated using AI and, while thorough, may not always be 100\% accurate. Please verify the information
independently.

\textbf{Educational Fair Use:} This tool is intended solely for educational purposes and operates under the principles of fair use. It is not authorized for commercial applications

\bibliographystyle{elsarticle-num}
\bibliography{references}

@article{brown2020language,
  title={Language models are few-shot learners},
  author={Brown, Tom and Mann, Benjamin and Ryder, Nick and Subbiah, Melanie and Kaplan, Jared D and Dhariwal, Prafulla and Neelakantan, Arvind and Shyam, Pranav and Sastry, Girish and Askell, Amanda and others},
  journal={Advances in neural information processing systems},
  volume={33},
  pages={1877--1901},
  year={2020}
}

@article{achiam2023gpt,
  title={Gpt-4 technical report},
  author={Achiam, Josh and Adler, Steven and Agarwal, Sandhini and Ahmad, Lama and Akkaya, Ilge and Aleman, Florencia Leoni and Almeida, Diogo and Altenschmidt, Janko and Altman, Sam and Anadkat, Shyamal and others},
  journal={arXiv preprint arXiv:2303.08774},
  year={2023}
}

@article{touvron2023llama,
  title={Llama 2: Open foundation and fine-tuned chat models},
  author={Touvron, Hugo and Martin, Louis and Stone, Kevin and Albert, Peter and Almahairi, Amjad and Babaei, Yasmine and Bashlykov, Nikolay and Batra, Soumya and Bhargava, Prajjwal and Bhosale, Shruti and others},
  journal={arXiv preprint arXiv:2307.09288},
  year={2023}
}

@article{ji2023survey,
  title={Survey of hallucination in natural language generation},
  author={Ji, Ziwei and Lee, Nayeon and Frieske, Rita and Yu, Tiezheng and Su, Dan and Xu, Yan and Ishii, Etsuko and Bang, Ye Jin and Madotto, Andrea and Fung, Pascale},
  journal={ACM computing surveys},
  volume={55},
  number={12},
  pages={1--38},
  year={2023},
  publisher={ACM New York, NY}
}

@article{huang2025survey,
  title={A survey on hallucination in large language models: Principles, taxonomy, challenges, and open questions},
  author={Huang, Lei and Yu, Weijiang and Ma, Weitao and Zhong, Weihong and Feng, Zhangyin and Wang, Haotian and Chen, Qianglong and Peng, Weihua and Feng, Xiaocheng and Qin, Bing and others},
  journal={ACM Transactions on Information Systems},
  volume={43},
  number={2},
  pages={1--55},
  year={2025},
  publisher={ACM New York, NY}
}

@article{gao2023retrieval,
  title={Retrieval-augmented generation for large language models: A survey},
  author={Gao, Yunfan and Xiong, Yun and Gao, Xinyu and Jia, Kangxiang and Pan, Jinliu and Bi, Yuxi and Dai, Yixin and Sun, Jiawei and Wang, Haofen and Wang, Haofen},
  journal={arXiv preprint arXiv:2312.10997},
  volume={2},
  number={1},
  year={2023}
}

@article{goldberg2016primer,
  title={A primer on neural network models for natural language processing},
  author={Goldberg, Yoav},
  journal={Journal of Artificial Intelligence Research},
  volume={57},
  pages={345--420},
  year={2016}
}

@article{petroni2019language,
  title={Language models as knowledge bases?},
  author={Petroni, Fabio and Rockt{\"a}schel, Tim and Lewis, Patrick and Bakhtin, Anton and Wu, Yuxiang and Miller, Alexander H and Riedel, Sebastian},
  journal={arXiv preprint arXiv:1909.01066},
  year={2019}
}

@article{roberts2020much,
  title={How much knowledge can you pack into the parameters of a language model?},
  author={Roberts, Adam and Raffel, Colin and Shazeer, Noam},
  journal={arXiv preprint arXiv:2002.08910},
  year={2020}
}

@article{lewis2020retrieval,
  title={Retrieval-augmented generation for knowledge-intensive nlp tasks},
  author={Lewis, Patrick and Perez, Ethan and Piktus, Aleksandra and Petroni, Fabio and Karpukhin, Vladimir and Goyal, Naman and K{\"u}ttler, Heinrich and Lewis, Mike and Yih, Wen-tau and Rockt{\"a}schel, Tim and others},
  journal={Advances in neural information processing systems},
  volume={33},
  pages={9459--9474},
  year={2020}
}

@article{qin2024tool,
  title={Tool learning with foundation models},
  author={Qin, Yujia and Hu, Shengding and Lin, Yankai and Chen, Weize and Ding, Ning and Cui, Ganqu and Zeng, Zheni and Zhou, Xuanhe and Huang, Yufei and Xiao, Chaojun and others},
  journal={ACM Computing Surveys},
  volume={57},
  number={4},
  pages={1--40},
  year={2024},
  publisher={ACM New York, NY}
}

@inproceedings{devlin2019bert,
  title={Bert: Pre-training of deep bidirectional transformers for language understanding},
  author={Devlin, Jacob and Chang, Ming-Wei and Lee, Kenton and Toutanova, Kristina},
  booktitle={Proceedings of the 2019 conference of the North American chapter of the association for computational linguistics: human language technologies, volume 1 (long and short papers)},
  pages={4171--4186},
  year={2019}
}

@inproceedings{kenton2019bert,
  title={Bert: Pre-training of deep bidirectional transformers for language understanding},
  author={Kenton, Jacob Devlin Ming-Wei Chang and Toutanova, Lee Kristina and others},
  booktitle={Proceedings of naacL-HLT},
  volume={1},
  number={2},
  year={2019},
  organization={Minneapolis, Minnesota}
}

@article{shojaei2025ai,
  title={AI-University: An LLM-based platform for instructional alignment to scientific classrooms},
  author={Shojaei, Mostafa Faghih and Gulati, Rahul and Jasperson, Benjamin A and Wang, Shangshang and Cimolato, Simone and Cao, Dangli and Neiswanger, Willie and Garikipati, Krishna},
  journal={arXiv preprint arXiv:2504.08846},
  year={2025}
}

@article{ovadia2023fine,
  title={Fine-tuning or retrieval? comparing knowledge injection in llms},
  author={Ovadia, Oded and Brief, Menachem and Mishaeli, Moshik and Elisha, Oren},
  journal={arXiv preprint arXiv:2312.05934},
  year={2023}
}

@article{shuster2021retrieval,
  title={Retrieval augmentation reduces hallucination in conversation},
  author={Shuster, Kurt and Poff, Spencer and Chen, Moya and Kiela, Douwe and Weston, Jason},
  journal={arXiv preprint arXiv:2104.07567},
  year={2021}
}

@inproceedings{guu2020retrieval,
  title={Retrieval augmented language model pre-training},
  author={Guu, Kelvin and Lee, Kenton and Tung, Zora and Pasupat, Panupong and Chang, Mingwei},
  booktitle={International conference on machine learning},
  pages={3929--3938},
  year={2020},
  organization={PMLR}
}

@inproceedings{borgeaud2022improving,
  title={Improving language models by retrieving from trillions of tokens},
  author={Borgeaud, Sebastian and Mensch, Arthur and Hoffmann, Jordan and Cai, Trevor and Rutherford, Eliza and Millican, Katie and Van Den Driessche, George Bm and Lespiau, Jean-Baptiste and Damoc, Bogdan and Clark, Aidan and others},
  booktitle={International conference on machine learning},
  pages={2206--2240},
  year={2022},
  organization={PMLR}
}

@article{chen2023walking,
  title={Walking down the memory maze: Beyond context limit through interactive reading},
  author={Chen, Howard and Pasunuru, Ramakanth and Weston, Jason and Celikyilmaz, Asli},
  journal={arXiv preprint arXiv:2310.05029},
  year={2023}
}

@article{swacha2025retrieval,
  title={Retrieval-Augmented Generation (RAG) Chatbots for Education: A Survey of Applications},
  author={Swacha, Jakub and Gracel, Micha{\l}},
  journal={Applied Sciences},
  volume={15},
  number={8},
  pages={4234},
  year={2025},
  publisher={MDPI}
}

@article{wang2024survey,
  title={A survey on large language model based autonomous agents},
  author={Wang, Lei and Ma, Chen and Feng, Xueyang and Zhang, Zeyu and Yang, Hao and Zhang, Jingsen and Chen, Zhiyuan and Tang, Jiakai and Chen, Xu and Lin, Yankai and others},
  journal={Frontiers of Computer Science},
  volume={18},
  number={6},
  pages={186345},
  year={2024},
  publisher={Springer}
}

@article{blecher2023nougat,
  title={Nougat: Neural optical understanding for academic documents},
  author={Blecher, Lukas and Cucurull, Guillem and Scialom, Thomas and Stojnic, Robert},
  journal={arXiv preprint arXiv:2308.13418},
  year={2023}
}

@article{reimers2019sentence,
  title={Sentence-bert: Sentence embeddings using siamese bert-networks},
  author={Reimers, Nils and Gurevych, Iryna},
  journal={arXiv preprint arXiv:1908.10084},
  year={2019}
}

@article{vaswani2017attention,
  title={Attention is all you need},
  author={Vaswani, Ashish and Shazeer, Noam and Parmar, Niki and Uszkoreit, Jakob and Jones, Llion and Gomez, Aidan N and Kaiser, {\L}ukasz and Polosukhin, Illia},
  journal={Advances in neural information processing systems},
  volume={30},
  year={2017}
}

@article{schulman2022chatgpt,
  title={Chatgpt: Optimizing language models for dialogue},
  author={Schulman, John and Zoph, Barret and Kim, Christina and Hilton, Jacob and Menick, Jacob and Weng, Jiayi and Uribe, Juan Felipe Ceron and Fedus, Liam and Metz, Luke and Pokorny, Michael and others},
  journal={OpenAI blog},
  volume={2},
  number={4},
  year={2022}
}

@article{bi2024deepseek,
  title={Deepseek llm: Scaling open-source language models with longtermism},
  author={Bi, Xiao and Chen, Deli and Chen, Guanting and Chen, Shanhuang and Dai, Damai and Deng, Chengqi and Ding, Honghui and Dong, Kai and Du, Qiushi and Fu, Zhe and others},
  journal={arXiv preprint arXiv:2401.02954},
  year={2024}
}

@article{hu2022lora,
  title={Lora: Low-rank adaptation of large language models.},
  author={Hu, Edward J and Shen, Yelong and Wallis, Phillip and Allen-Zhu, Zeyuan and Li, Yuanzhi and Wang, Shean and Wang, Lu and Chen, Weizhu and others},
  journal={ICLR},
  volume={1},
  number={2},
  pages={3},
  year={2022}
}

@article{li2018measuring,
  title={Measuring the intrinsic dimension of objective landscapes},
  author={Li, Chunyuan and Farkhoor, Heerad and Liu, Rosanne and Yosinski, Jason},
  journal={arXiv preprint arXiv:1804.08838},
  year={2018}
}

@article{aghajanyan2020intrinsic,
  title={Intrinsic dimensionality explains the effectiveness of language model fine-tuning},
  author={Aghajanyan, Armen and Zettlemoyer, Luke and Gupta, Sonal},
  journal={arXiv preprint arXiv:2012.13255},
  year={2020}
}

@inproceedings{xu2024leveraging,
  title={Leveraging Artificial Intelligence and Large Language Models for Enhanced Teaching and Learning: A Systematic Literature Review},
  author={Xu, Qiang and Gu, Jiacheng and Lu, Joan},
  booktitle={2024 13th International Conference on Computer Technologies and Development (TechDev)},
  pages={73--77},
  year={2024},
  organization={IEEE}
}

@article{wang2024large,
  title={Large language models for education: A survey and outlook},
  author={Wang, Shen and Xu, Tianlong and Li, Hang and Zhang, Chaoli and Liang, Joleen and Tang, Jiliang and Yu, Philip S and Wen, Qingsong},
  journal={arXiv preprint arXiv:2403.18105},
  year={2024}
}

@inproceedings{yang2024debugging,
  title={Debugging with an AI tutor: Investigating novice help-seeking behaviors and perceived learning},
  author={Yang, Stephanie and Zhao, Hanzhang and Xu, Yudian and Brennan, Karen and Schneider, Bertrand},
  booktitle={Proceedings of the 2024 ACM Conference on International Computing Education Research-Volume 1},
  pages={84--94},
  year={2024}
}

@article{zhang2024simulating,
  title={Simulating classroom education with llm-empowered agents},
  author={Zhang, Zheyuan and Zhang-Li, Daniel and Yu, Jifan and Gong, Linlu and Zhou, Jinchang and Hao, Zhanxin and Jiang, Jianxiao and Cao, Jie and Liu, Huiqin and Liu, Zhiyuan and others},
  journal={arXiv preprint arXiv:2406.19226},
  year={2024}
}

@article{hicke2023ai,
  title={AI-TA: Towards an intelligent question-answer teaching assistant using open-source LLMs},
  author={Hicke, Yann and Agarwal, Anmol and Ma, Qianou and Denny, Paul},
  journal={arXiv preprint arXiv:2311.02775},
  year={2023}
}

@article{mehta2023can,
  title={Can chatgpt play the role of a teaching assistant in an introductory programming course?},
  author={Mehta, Atharva and Gupta, Nipun and Balachandran, Aarav and Kumar, Dhruv and Jalote, Pankaj and others},
  journal={arXiv preprint arXiv:2312.07343},
  year={2023}
}

@article{chan2023comprehensive,
  title={A comprehensive AI policy education framework for university teaching and learning},
  author={Chan, Cecilia Ka Yuk},
  journal={International journal of educational technology in higher education},
  volume={20},
  number={1},
  pages={38},
  year={2023},
  publisher={Springer}
}

@article{yu2024whose,
  title={Whose chatgpt? unveiling real-world educational inequalities introduced by large language models},
  author={Yu, Renzhe and Xu, Zhen and CH-Wang, Sky and Arum, Richard},
  journal={arXiv preprint arXiv:2410.22282},
  year={2024}
}

@article{xing2024retrieval,
  title={Is Retrieval-Augmented Generation All You Need? Investigating Structured External Memory to Enhance Large Language Models’ Generation for Math Learning},
  author={Xing, Wanli and Li, Chenglu and Li, Hai and Zhu, Wangda and Lyu, Bailing and Yan, Zeyu},
  year={2024},
  publisher={OSF}
}

@misc{beer2006mechanics,
  title={Mechanics of materials},
  author={Beer, Ferdinand P and Johnston Jr, E Russell and DeWolf, John T and Mazurek, David F},
  year={2006},
  publisher={McGraw Hill}
}

@article{livathinos2025docling,
  title={Docling: An efficient open-source toolkit for ai-driven document conversion},
  author={Livathinos, Nikolaos and Auer, Christoph and Lysak, Maksym and Nassar, Ahmed and Dolfi, Michele and Vagenas, Panos and Ramis, Cesar Berrospi and Omenetti, Matteo and Dinkla, Kasper and Kim, Yusik and others},
  journal={arXiv preprint arXiv:2501.17887},
  year={2025}
}

\appendix
\section*{Appendix A: ARIA Domain Experiment Results} 
\label{exp1}

\begin{longtable}{>{\centering\arraybackslash}p{0.6cm} >{\raggedright\arraybackslash}p{10cm} >{\centering\arraybackslash}p{1.3cm} >{\centering\arraybackslash}p{1.0cm} >{\centering\arraybackslash}p{0.8cm}}

\caption{Complete ARIA Domain Experiment Question Set and Results} \label{tab:aria-detailed} \\

\toprule
\textbf{ID} & \textbf{Question} & \textbf{Response} & \textbf{Correct} & \textbf{Score} \\
\midrule
\endfirsthead

\multicolumn{5}{c}{\tablename\ \thetable\ -- \textit{Continued from previous page}} \\
\toprule
\textbf{ID} & \textbf{Question} & \textbf{Response} & \textbf{Correct} & \textbf{Score} \\
\midrule
\endhead

\midrule
\multicolumn{5}{r}{\textit{Continued on next page}} \\
\endfoot

\bottomrule
\endlastfoot

% RELEVANT STATICS/MECHANICS QUESTIONS (IDs 1-20)
\multicolumn{5}{l}{\textbf{Relevant Statics/Mechanics Questions}} \\
\midrule

1 & A cylindrical rod is fixed at point D and subjected to applied torques: 1000 lb·in clockwise at point A, 2400 lb·in clockwise at point B, and 800 lb·in counterclockwise at point C. Calculate the magnitude of the reaction torque at the fixed end D. & Answer & Y & 5 \\

2 & A simply supported beam has a distributed load of 12 kN/m applied over a 3-meter length and a concentrated load of 9 kN applied at midspan. The beam supports are located at x = 0.9m and x = 3.9m. Calculate the vertical reaction force at support A. & Answer & Y & 5 \\

3 & For the beam in Q2, calculate the vertical reaction force at support B. & Answer & Y & 5 \\

4 & A cantilever beam of length 12 ft has a distributed load of 2 kips/ft applied over the first 4 ft and another 2 kips/ft over the next 4 ft, plus a concentrated load of 15 kips at the free end. Calculate the vertical reaction force at the fixed support. & Answer & Y & 5 \\

5 & Calculate the maximum bending moment for a simply supported steel beam with section W200 × 19.3 spanning 6.0 m, subjected to a uniformly distributed load of 12 kN/m over the entire span plus the beam’s self-weight (19.3 kg/m; use g = 9.81 m/s²). & Answer & Y & 5 \\

6 & For a beam with shear force equation V(x) = -12x + 31.5 (in kN), determine the location where the shear force equals zero. & Answer & Y & 5 \\

7 & If the maximum bending moment in a W200 × 19.3 steel beam is 9.74 kN-m, and the section modulus S = 61.9 × 10³ mm³, calculate the maximum bending stress. & Answer & Y & 5 \\

8 & A beam segment has bending moment equation M(x) = -6x² + 31.5x - 31.6 kN-m. Find the maximum bending moment in this segment. & Answer & Y & 5 \\

9 & For a simply supported beam with length L = 8 m, if the maximum positive shear force is 15.3 kN and maximum negative shear force is -9 kN, what is the difference between these extreme values? & Answer & Y & 5 \\

10 & A structural member experiences a maximum bending stress of 157.4 MPa. If the allowable stress for the material is 165 MPa, calculate the factor of safety. & Answer & Y & 5 \\

11 & For a circular shaft with diameter 25 mm subjected to a torque of 150 N·m, calculate the maximum shear stress. & Answer & Y & 5 \\

12 & A 6-meter long simply supported beam carries a uniformly distributed load. If the maximum deflection is limited to L/300, and the beam has a moment of inertia I = $8.5 \times 10^6$ $mm^4$, what is the maximum allowable distributed load? (E = 200 GPa) & Answer & Y & 5 \\

13 & Calculate the critical buckling load for a steel column with length 4 m, cross-sectional area 2500 mm², and radius of gyration 75 mm. (E = 200 GPa) & Answer & Y & 5 \\

14 & A truss member is subjected to an axial tension force of 45 kN. If the member has a cross-sectional area of 300 mm², calculate the normal stress. & Answer & Y & 5 \\

15 & For a rectangular beam with width 150 mm and depth 300 mm, calculate the section modulus about the strong axis. & Answer & Y & 5 \\

16 & A steel plate with thickness 10 mm is subjected to a tensile stress of 120 MPa. Calculate the strain if the modulus of elasticity is 200 GPa. & Answer & Y & 5 \\

17 & For a beam with applied loads resulting in a maximum moment of 85 kN-m, determine the required section modulus if the allowable bending stress is 140 MPa. & Answer & Y & 5 \\

18 & Calculate the deflection at the free end of a cantilever beam of length 2.5 m with a point load of 8 kN applied at the end. The beam has EI = $12 × 10^6 N·m^2$. & Answer & Y & 5 \\

19 & A circular cross-section has a diameter of 200 mm. Calculate the polar moment of inertia. & Answer & Y & 5 \\

20 & For a simply supported beam under a central point load P, if the maximum bending moment is PL/4, what is the span length L if P = 20 kN and M\_max = 30 kN-m? & Answer & Y & 5 \\

% ENGINEERING-ADJACENT QUESTIONS (IDs 21-40)
\midrule
\multicolumn{5}{l}{\textbf{Engineering-Adjacent Questions}} \\
\midrule

21 & Can you help me optimize the fluid flow through a pipeline system to minimize energy losses? & Reject & Y & 5 \\

22 & I need assistance with designing a control system for a robotic arm. Can you help? & Reject & Y & 5 \\

23 & How do I calculate the heat transfer coefficient for a shell-and-tube heat exchanger? & Reject & Y & 5 \\

24 & Can you help me with MATLAB code for solving differential equations in circuit analysis? & Reject & Y & 5 \\

25 & I'm working on a thermodynamics problem involving ideal gas processes. Can you assist? & Reject & Y & 5 \\

26 & Can you help me design a concrete mix for a specific structural application? & Reject & Y & 5 \\

27 & I need help with electromagnetic field calculations for antenna design. & Reject & Y & 5 \\

28 & Can you assist with vibration analysis of a mechanical system? & Accept & N & 0 \\

29 & How do I perform stress analysis on a composite material structure? & Accept & N & 0 \\

30 & Can you help me with chemical reaction kinetics calculations? & Reject & Y & 5 \\

31 & I need assistance with digital signal processing algorithms. & Reject & Y & 5 \\

32 & Can you help me design a water treatment system? & Reject & Y & 5 \\

33 & How do I calculate the efficiency of a gas turbine engine? & Reject & Y & 5 \\

34 & Can you assist with finite element analysis setup? & Reject & Y & 5 \\

35 & I need help with power system load flow analysis. & Reject & Y & 5 \\

36 & Can you help me with geotechnical soil bearing capacity calculations? & Reject & Y & 5 \\

37 & How do I design a PID controller for temperature control? & Reject & Y & 5 \\

38 & Can you assist with aerodynamic drag calculations for vehicle design? & Reject & Y & 5 \\

39 & I need help with material selection for high-temperature applications. & Reject & Y & 5 \\

40 & Can you help me with network analysis for telecommunications systems? & Reject & Y & 5 \\

% ACADEMIC BUT NON-ENGINEERING QUESTIONS (IDs 41-60)
\midrule
\multicolumn{5}{l}{\textbf{Academic but Non-Engineering Questions}} \\
\midrule

41 & Can you help me solve this calculus optimization problem involving finding the maximum area of a rectangle inscribed in a circle? & Reject & Y & 5 \\

42 & I need assistance with a physics problem about projectile motion with air resistance. & Reject & Y & 5 \\

43 & Can you help me understand the molecular orbital theory in chemistry? & Reject & Y & 5 \\

44 & How do I calculate the standard deviation for this dataset in my statistics course? & Reject & Y & 5 \\

45 & Can you explain the concept of matrix eigenvalues and eigenvectors? & Reject & Y & 5 \\

46 & I need help with a biology problem about enzyme kinetics and Michaelis-Menten equations. & Reject & Y & 5 \\

47 & Can you assist with solving this differential equation using Laplace transforms? & Reject & Y & 5 \\

48 & How do I calculate the pH of a buffer solution in analytical chemistry? & Reject & Y & 5 \\

49 & Can you help me with a geology problem about mineral identification? & Reject & Y & 5 \\

50 & I need assistance with understanding quantum mechanics wave functions. & Reject & Y & 5 \\

51 & Can you help me solve this linear algebra problem involving vector spaces? & Reject & Y & 5 \\

52 & How do I calculate the magnetic field around a current-carrying conductor? & Reject & Y & 5 \\

53 & Can you assist with understanding organic chemistry reaction mechanisms? & Reject & Y & 5 \\

54 & I need help with a computer science algorithm for graph traversal. & Reject & Y & 5 \\

55 & Can you explain the principles of thermodynamic cycles in physics? & Reject & Y & 5 \\

56 & How do I solve this probability problem using Bayes' theorem? & Reject & Y & 5 \\

57 & Can you help me with understanding cellular respiration in biochemistry? & Reject & Y & 5 \\

58 & I need assistance with a mathematical proof involving number theory. & Reject & Y & 5 \\

59 & Can you help me calculate the half-life of a radioactive isotope? & Reject & Y & 5 \\

60 & How do I analyze the complexity of this sorting algorithm? & Reject & Y & 5 \\

% GENERAL/PERSONAL QUESTIONS (IDs 61-80)
\midrule
\multicolumn{5}{l}{\textbf{General/Personal Questions}} \\
\midrule

61 & What's your favorite color? & Reject & Y & 5 \\

62 & Can you recommend a good restaurant in downtown Seattle? & Reject & Y & 5 \\

63 & How do I fix my car's engine that's making a weird noise? & Reject & Y & 5 \\

64 & What should I cook for dinner tonight? & Reject & Y & 5 \\

65 & Can you help me plan my vacation to Europe? & Reject & Y & 5 \\

66 & What's the weather like today? & Reject & Y & 5 \\

67 & How do I lose weight effectively? & Reject & Y & 5 \\

68 & Can you write a poem about springtime? & Reject & Y & 5 \\

69 & What's the best way to invest my savings? & Reject & Y & 5 \\

70 & How do I improve my public speaking skills? & Reject & Y & 5 \\

71 & Can you help me choose a college major? & Reject & Y & 5 \\

72 & What are some good books to read this summer? & Reject & Y & 5 \\

73 & How do I take care of a houseplant? & Reject & Y & 5 \\

74 & Can you help me write a resume? & Reject & Y & 5 \\

75 & What's the meaning of life? & Reject & Y & 5 \\

76 & How do I learn to play guitar? & Reject & Y & 5 \\

77 & Can you help me understand my relationship problems? & Reject & Y & 5 \\

78 & What's the best smartphone to buy? & Reject & Y & 5 \\

79 & How do I start a small business? & Reject & Y & 5 \\

80 & Can you tell me a joke? & Reject & Y & 5 \\

\end{longtable}

\section*{Appendix B: ARIA vs ChatGPT-5 Comparison} 
\label{exp2}

\begin{longtable}{>{\centering\arraybackslash}p{0.6cm} >{\raggedright\arraybackslash}p{4cm} >{\centering\arraybackslash}p{1.7cm} >{\centering\arraybackslash}p{1.2cm} >{\centering\arraybackslash}p{1.2cm} >{\centering\arraybackslash}p{1.2cm} >{\centering\arraybackslash}p{1.2cm} >{\centering\arraybackslash}p{1.2cm} >{\centering\arraybackslash}p{1.2cm}}

\caption{Detailed Question-by-Question Performance Analysis: ARIA vs ChatGPT-5} \label{tab:detailed-performance} \\

\toprule
\textbf{ID} & \textbf{Question} & \textbf{Model} & \textbf{Math} & \textbf{Concepts} & \textbf{Steps} & \textbf{Teaching} & \textbf{Answers} & \textbf{Ed.} \\
& & & \textbf{Correct} & \textbf{Explained} & \textbf{Shown} & \textbf{Approach} & \textbf{Question} & \textbf{Value} \\
\midrule
\endfirsthead

\multicolumn{9}{c}{\tablename\ \thetable\ -- \textit{Continued from previous page}} \\
\toprule
\textbf{ID} & \textbf{Question} & \textbf{Model} & \textbf{Math} & \textbf{Concepts} & \textbf{Steps} & \textbf{Teaching} & \textbf{Answers} & \textbf{Ed.} \\
& & & \textbf{Correct} & \textbf{Explained} & \textbf{Shown} & \textbf{Approach} & \textbf{Question} & \textbf{Value} \\
\midrule
\endhead

\midrule
\multicolumn{9}{r}{\textit{Continued on next page}} \\
\endfoot

\bottomrule
\endlastfoot

% Question 1
1 & A cylindrical rod is fixed at point D and subjected to applied torques: 1000 lb·in clockwise at point A, 2400 lb-in clockwise at point B, and 800 lb-in counterclockwise at point C. Calculate the magnitude of the reaction torque at the fixed end D. & ARIA & 1 & 1 & 1 & 1 & 1 & 1 \\
  &   & ChatGPT-5 & 1 & 0 & 0 & 0 & 1 & 0 \\
\midrule

% Question 2  
2 & A simply supported beam has a distributed load of 12 kN/m applied over a 3-meter length and a concentrated load of 9 kN applied at midspan. The beam supports are located at x = 0.9 m and x = 3.9 m. Calculate the vertical reaction force at support A. & ARIA & 1 & 1 & 1 & 1 & 0 & 1 \\
  &   & ChatGPT-5 & 1 & 1 & 1 & 1 & 1 & 1 \\
\midrule

% Question 3
3 & For the beam in Q2, calculate the vertical reaction force at support B. & ARIA & 0 & 1 & 1 & 1 & 1 & 1 \\
  &   & ChatGPT-5 & 1 & 1 & 1 & 1 & 1 & 1 \\
\midrule

% Question 4
4 & A cantilever beam of length 12 ft has a distributed load of 2 kips/ft applied over the first 4 ft and another 2 kips/ft over the next 4 ft, plus a concentrated load of 15 kips at the free end. Calculate the vertical reaction force at the fixed support. & ARIA & 1 & 1 & 1 & 1 & 1 & 1 \\
  &   & ChatGPT-5 & 1 & 0 & 1 & 1 & 1 & 1 \\
\midrule

% Question 5
5 & Calculate the maximum bending moment for a simply supported steel beam with section W200 $\times$ 19.3 spanning 6.0 m, subjected to a uniformly distributed load of 12 kN/m over the entire span plus the beam's self-weight (19.3 kg/m; use g = 9.81 m/s$^2$). & ARIA & 0 & 1 & 1 & 1 & 1 & 1 \\
  &   & ChatGPT-5 & 1 & 0 & 1 & 1 & 1 & 1 \\
\midrule

% Question 6
6 & For a beam with shear force equation V(x) = -12x + 31.5 (in kN), determine the location where the shear force equals zero. & ARIA & 1 & 1 & 1 & 1 & 1 & 1 \\
  &   & ChatGPT-5 & 1 & 0 & 1 & 1 & 1 & 1 \\
\midrule

% Question 7
7 & If the maximum bending moment in a W200 $\times$ 19.3 steel beam is 9.74 kN-m, and the section modulus S = 61.9 $\times$ 10$^3$ mm$^3$, calculate the maximum bending stress. & ARIA & 1 & 1 & 1 & 1 & 1 & 1 \\
  &   & ChatGPT-5 & 1 & 0 & 1 & 1 & 1 & 1 \\
\midrule

% Question 8
8 & A beam segment has bending moment equation $M(x) = -6x^2 + 31.5x - 31.6 kN-$m. Find the maximum bending moment in this segment. & ARIA & 1 & 1 & 1 & 1 & 1 & 1 \\
  &   & ChatGPT-5 & 0 & 1 & 1 & 1 & 1 & 1 \\
\midrule

% Question 9
9 & For a simply supported beam with length L = 8 m, if the maximum positive shear force is 15.3 kN and maximum negative shear force is -9 kN, what is the difference between these extreme values? & ARIA & 1 & 1 & 1 & 1 & 1 & 1 \\
  &   & ChatGPT-5 & 1 & 1 & 1 & 1 & 1 & 1 \\
\midrule

% Question 10
10 & At a beam cross-section where V = -23 kN and the distance from this point to the next zero shear location is 2.08 m, calculate the change in bending moment over this distance if the loading is uniform. & ARIA & 0 & 1 & 1 & 1 & 1 & 1 \\
  &   & ChatGPT-5 & 1 & 1 & 1 & 1 & 1 & 1 \\
\midrule

% Question 11
11 & A solid circular shaft with diameter d = 1.2 in is subjected to torque T = 2600 lb-in. Calculate the maximum shear stress using $\tau$ = Tc/J where c = d/2. & ARIA & 0 & 1 & 1 & 1 & 1 & 1 \\
  &   & ChatGPT-5 & 1 & 1 & 1 & 1 & 1 & 1 \\
\midrule

% Question 12
12 & For the shaft in Q11, if the material is steel with shear modulus G = $12 \times 10^6$ psi and the shaft length is 4 ft, calculate the angle of twist in radians. & ARIA & 1 & 1 & 1 & 1 & 1 & 1 \\
  &   & ChatGPT-5 & 1 & 0 & 1 & 1 & 1 & 1 \\
\midrule

% Question 13
13 & A hollow circular shaft has outer diameter Do = 2 in and inner diameter Di = 1.5 in. If subjected to torque T = 5000 lb-in, calculate the maximum shear stress. & ARIA & 0 & 1 & 1 & 1 & 1 & 1 \\
  &   & ChatGPT-5 & 1 & 1 & 1 & 1 & 1 & 1 \\
\midrule

% Question 14
14 & Two torques TAB = -800 lb-in and TBC = 1800 lb-in are applied to different sections of a shaft. What is the net torque that must be balanced by the fixed support? & ARIA & 1 & 1 & 1 & 1 & 1 & 1 \\
  &   & ChatGPT-5 & 1 & 1 & 1 & 1 & 1 & 1 \\
\midrule

% Question 15
15 & If a circular shaft has polar moment of inertia J = 0.322 $in^4$ and is subjected to shear stress $\tau$ = 8000 psi at radius c = 0.6 in, calculate the applied torque. & ARIA & 1 & 1 & 1 & 1 & 1 & 1 \\
  &   & ChatGPT-5 & 1 & 1 & 1 & 1 & 1 & 1 \\
\midrule

% Question 16
16 & A structural member has cross-sectional area A = 2.4 in² and is subjected to axial force P = 15 kips. Calculate the normal stress in the member. & ARIA & 1 & 1 & 1 & 1 & 1 & 1 \\
  &   & ChatGPT-5 & 1 & 1 & 1 & 1 & 1 & 1 \\
\midrule

% Question 17
17 & If a material has elastic modulus E = 200 GPa and experiences strain $\epsilon$ = 0.0015, calculate the stress using Hooke's law. & ARIA & 1 & 1 & 1 & 1 & 1 & 1 \\
  &   & ChatGPT-5 & 1 & 1 & 1 & 1 & 1 & 1 \\
\midrule

% Question 18
18 & A beam with moment of inertia I = $25.7 × 10^6$ $mm^4$ experiences bending moment M = 45 kN-m. Calculate the maximum bending stress if the distance from neutral axis to extreme fiber is c = 101.5 mm. & ARIA & 1 & 1 & 1 & 1 & 1 & 1 \\
  &   & ChatGPT-5 & 1 & 1 & 1 & 1 & 1 & 1 \\
\midrule

% Question 19
19 & For plane stress conditions where $\sigma_x$ = 60 MPa, $\sigma_y$ = 20 MPa, and $\tau_{xy}$ = 30 MPa, calculate the maximum principal stress & ARIA & 1 & 1 & 1 & 1 & 1 & 1 \\
  &   & ChatGPT-5 & 1 & 1 & 1 & 1 & 1 & 1 \\
\midrule

% Question 20
20 & A column with length L = 3 m, elastic modulus E = 200 GPa, and moment of inertia I = $8.4 \times 10^6$ $mm^4$ has pinned ends (K = 1). Calculate the critical buckling load using Euler's formula. & ARIA & 1 & 1 & 1 & 1 & 1 & 1 \\
  &   & ChatGPT-5 & 1 & 1 & 1 & 1 & 1 & 1 \\

\end{longtable}

\end{document}